\documentclass[trackchanges]{aastex701}

\def \siiv {Si\,{\sc iv}}

\begin{document}

\title{Evidence for anisotropic non-Maxwellian velocity distributions in the solar transition region}

\author[0000-0002-0405-0668]{Paola Testa}
\affil{Center for Astrophysics | Harvard \& Smithsonian, 60 Garden Street, Cambridge, MA 02193, USA}
\email[show]{ptesta@cfa.harvard.edu}  

\author[0000-0002-8370-952X]{Bart De Pontieu}
\affil{Lockheed Martin Solar \& Astrophysics Laboratory, 3251 Hanover Street, Palo Alto, CA 94304, USA}
\affil{Rosseland Center for Solar Physics, University of Oslo, P.O. Box 1029 Blindern, NO-0315 Oslo, Norway}
\affil{Institute of Theoretical Astrophysics, University of Oslo, P.O. Box 1029 Blindern, NO-0315 Oslo, Norway}
\email{bdp@lmsal.com}  

\author[0000-0001-7460-725X]{Kyuhyoun Cho}
\affiliation{SETI Institute, 339 Bernardo Avenue, Suite 200 Mountain View, CA 94043, USA}
\affiliation{Lockheed Martin Solar \& Astrophysics Laboratory, 3251 Hanover Street, Palo Alto, CA 94304, USA}
\email{cho@seti.org}  

\author[0000-0002-1242-5124]{Thomas Ayres}
\affiliation{Center for Astrophysics and Space Astronomy, University of Colorado, Boulder, 80309, CO, USA}
\email{thomas.ayres@colorado.edu}  

\author[0000-0001-5690-2351]{Alexander J. B. Russell}
\affiliation{School of Mathematics \& Statistics, University of St Andrews, St Andrews, KY16 9SS, UK}
\email{ar51@st-andrews.ac.uk}

\begin{abstract}

Spectroscopic observations of the solar transition region have long shown excess line broadening and non-Gaussian profiles, but their spatial distribution and prevalence are not well established, and their physical origin remains the subject of debate. Here we analyze \siiv\ line profiles in full-disk mosaics of observations with Interface Region Imaging Spectrograph (IRIS; \citealt{DePontieu2014}), and find that non-Gaussian $\kappa$-like profiles, indicative of suprathermal velocity distributions, are pervasive (comprising $\sim$60\% of the observed profiles). In addition, we find that their occurrence depends strongly on the angle between the line of sight and the magnetic field as inferred from NLFFF extrapolations. $\kappa$-like profiles become increasingly prevalent when the magnetic field is oriented transverse to the observer, whereas Gaussian profiles become more prevalent when the field is aligned with the line of sight. This geometric dependence provides evidence that the observed velocity distributions in the transition region are non-Maxwellian and anisotropic. Our findings provide evidence for the importance, in the solar transition region, of kinetic processes, and the key role the magnetic field plays in the superposition of signals, and/or the generation or relaxation of non-Maxwellian particle distributions, likely caused by magnetic energy release through reconnection. These effects are not captured by the magnetohydrodynamic approaches that are typically invoked to study the solar atmosphere. Our results highlight the need and provide constraints for more advanced multi-fluid and/or kinetic modeling of the solar transition region.

\end{abstract}



\section{Introduction}\label{sec1}

The solar transition region, the narrow interface where the temperature rises dramatically from $\sim 10^4$~K in the chromosphere to over $10^6$~K in the corona, is highly sensitive to temperature, density, flows, and non-thermal motions.
Spectroscopy of the transition region emission lines provides uniquely powerful tools for probing the physics of the solar atmosphere because it provides direct, quantitative diagnostics of plasma conditions, allowing us to disentangle the complex interplay of heating, energy transport, and mass exchange in the solar atmosphere. 

The properties of line profiles of optically thin transition region emission lines provide diagnostics of waves and mass motions that transport energy into the corona, as these processes produce measurable effects such as non-thermal broadening and Doppler shifts, even when the observations lack spatial resolution.
Early Skylab observations already indicated that transition region line profiles, especially in network regions, can deviate from a single Gaussian and exhibit enhanced wings or multiple components \citep{Kjeldseth1977}.
Follow-up studies using observations from the sounding rocket HRTS have confirmed the presence of non-Gaussian profiles in active or network regions \citep{Dere1993}, also exploring these profiles in the context of bi-directional flows resulting from explosive events and small-scale reconnection \citep[e.g.,][]{Brueckner1983}.

\cite{Peter2000,Peter2001} carried out a more general study of non-Gaussian transition region profiles using SUMER spectra of quiet Sun, explored their spatial properties, and interpreted the wing enhancement in terms of a secondary component of a multi-Gaussian profile that is mostly present in bright quiet Sun network regions.

The higher spatial and spectral resolution of IRIS observations compared with previous spectrographs allows an accurate characterization of non-thermal widths and line  profiles for transition region emission \citep[e.g.,][]{DePontieu2015}. The non-Gaussian profiles observed with IRIS have recently been interpreted as $\kappa$ profiles arising from non-Maxwellian particle distributions \citep{Dudik2017}. 

Observational evidence for the presence of such non-thermal particles (e.g., accelerated electrons) outside of large flares has been obtained from hard X-ray observations for small (micro-) flares (e.g., \citealt{Hannah2008,Glesener2020,Polito2023,Testa2023book} and references therein). IRIS has provided further evidence of accelerated non-thermal particles for even smaller heating events, not easily accessible to hard X-ray observations \citep{Testa2014,Testa2020,Cho2023,Testa2023b}, suggesting the widespread presence of supra-thermal particles even in the quiescent corona.
For evidence of $\kappa$ profiles in solar flares, interpreted as evidence of non-thermal particles, see \citet{Jeffrey2016,Jeffrey2017} and \citet{Polito2018_kappa}.

\begin{figure}[!h]
\centering
\includegraphics[width=0.7\textwidth]{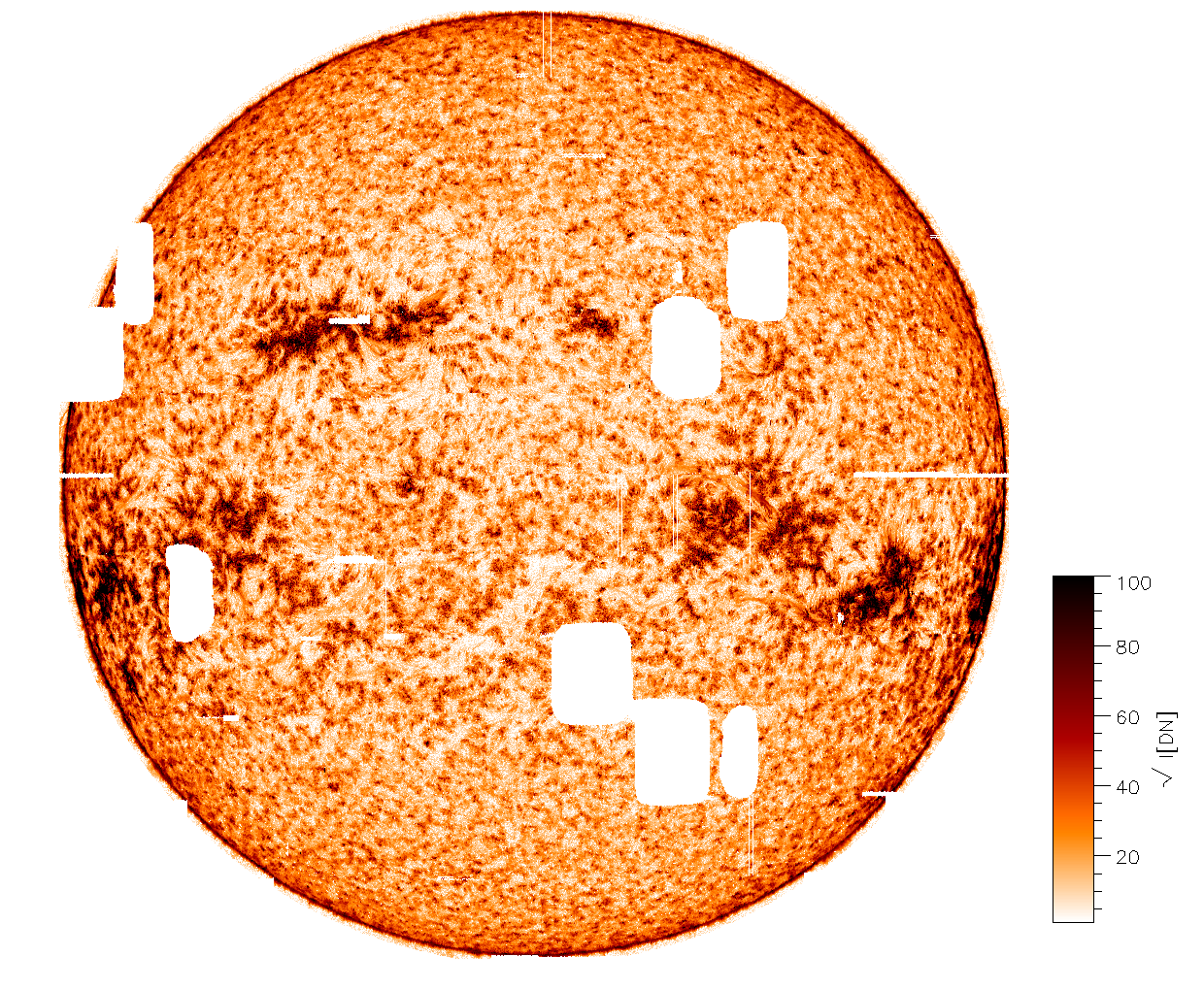}
\includegraphics[width=0.7\textwidth]{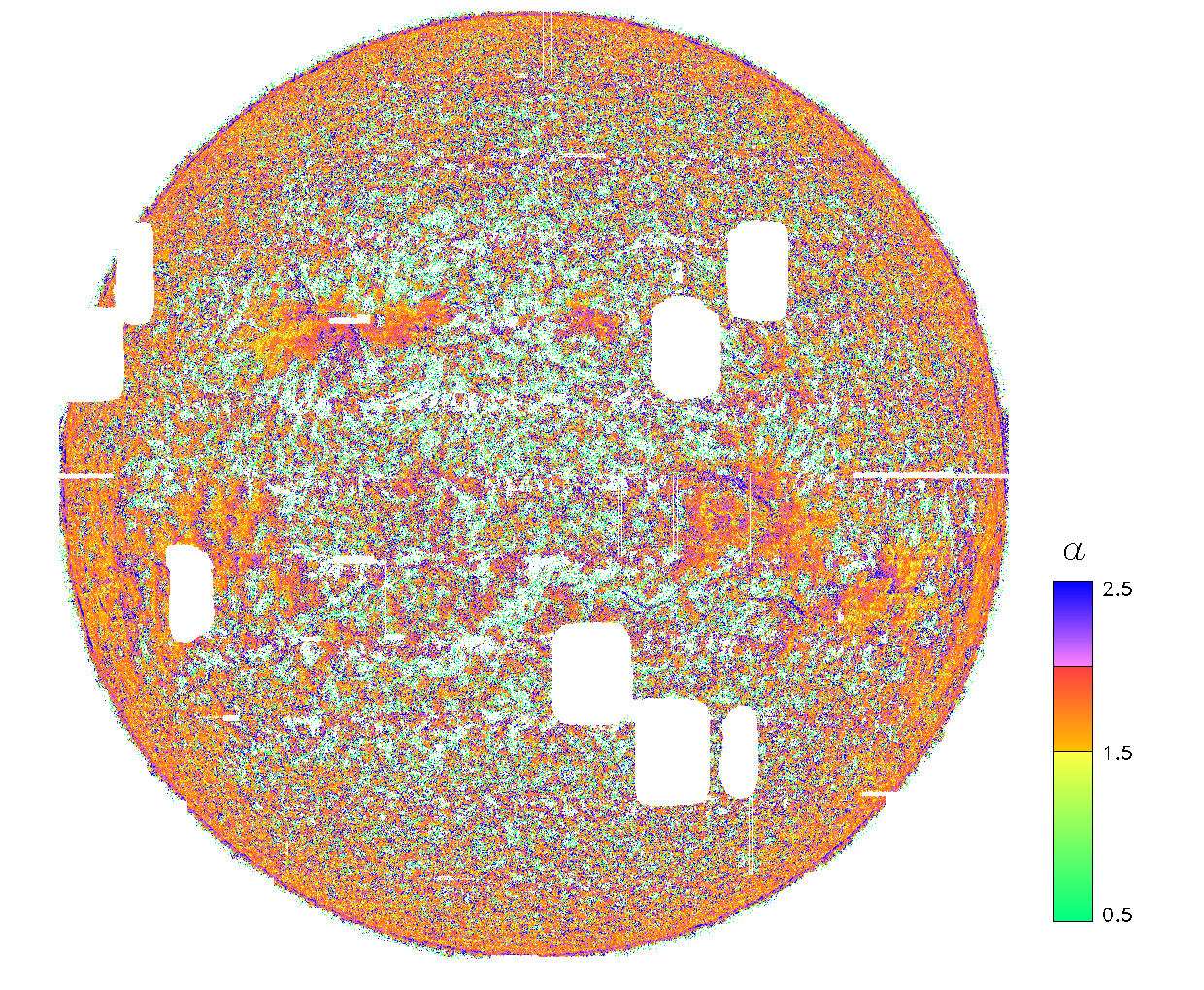}
\caption{\textbf{Maps of \siiv\ 1393.76\AA\ transition region line properties over the full solar disk, from IRIS full-disk mosaics of 2014-03-17.}
We show maps from pseudo-Gaussian line fitting for the \siiv\ 1393.76\AA\ line intensity (top) and $\alpha$ parameter describing the line shape (with $\alpha =2$, $< 2$ and $>2$ corresponding respectively to Gaussian profiles, $\kappa$-like profiles, and more rectangular profiles with truncated wings. White areas represent regions masked out from our analysis because they are affected by missing data or excessive particle
radiation (see \citealt{Ayres2021}). 
}\label{fig1}
\end{figure}

\cite{Ayres2021} used IRIS full-disk mosaics to investigate the properties of chromospheric and transition region lines in the context of the solar-stellar connection, and found for the \siiv\ 1393.76\AA\ transition line a common occurrence of non-Gaussian profiles showing for the first time how pervasive these non-Gaussian profiles are, down to the highest spatial resolution. Here we analyze one of the full-disk mosaics presented in \cite{Ayres2021} to investigate the relation of these non-Gaussian profiles with solar features and the local magnetic field direction, and the origin of such profiles (e.g., $\kappa$ distributions).

\begin{figure}[h!]
\centering
\includegraphics[width=0.9\textwidth]{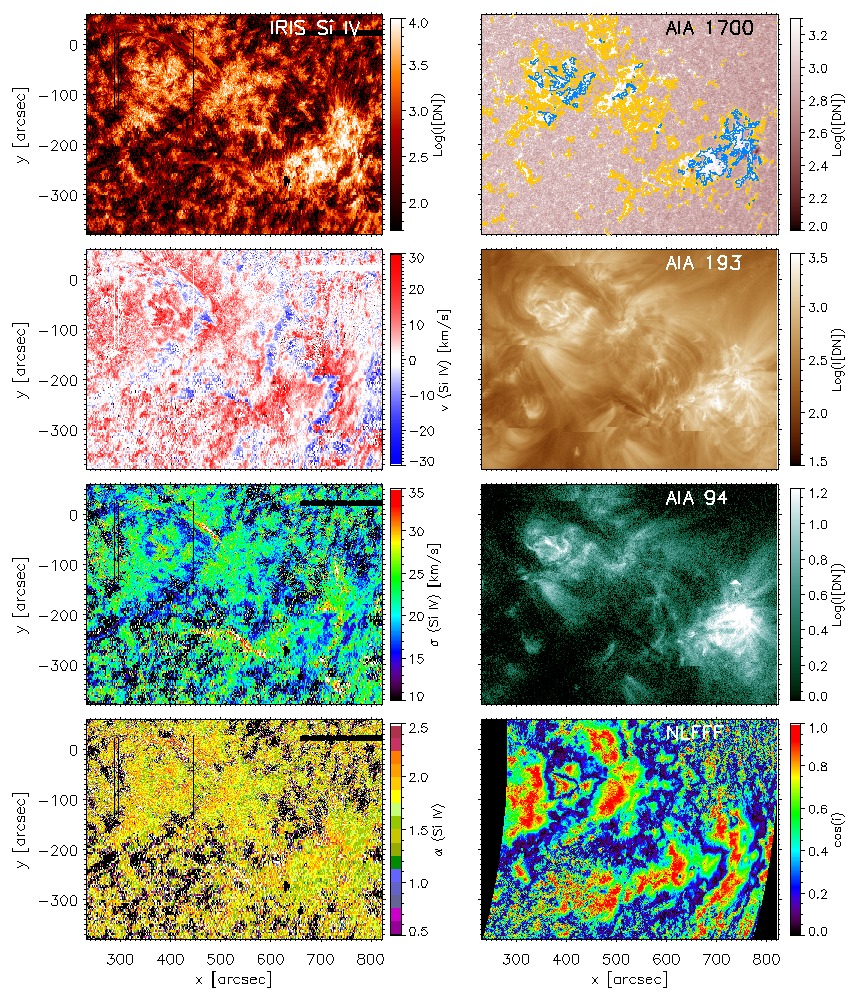}
\caption{\textbf{Transition region, coronal, photospheric, and magnetic field properties, derived from IRIS and SDO observations, for a region in the SW of the solar disk shown in Figure~\ref{fig1} including AR 12004 (at top left of the shown FOV) and AR 12002 (at bottom right, i.e., SW, corner).} Left column (top to bottom):  IRIS \siiv\ 1393.76\AA\ line intensity, Doppler shift (blue-shifts being due to motions toward the observer, and red-shifts to motions away from the observer), line width, and $\alpha$ parameter.  Right column (top to bottom): SDO/AIA 1700\AA\ (dominated by UV photospheric continuum) with contours delineating plage (yellow) and moss (blue) regions; AIA 193\AA\ and AIA 94\AA\ coronal images (with peak sensitivity to plasma at $\sim 1.5$~MK and  $8$~MK respectively); map of inclination between magnetic field (from Non-linear Force-Free Field, NLFFF, extrapolations) direction (at TR heights) and the line-of-sight.  See text for a detailed description of how these maps are derived. }\label{fig2}
\end{figure}

\section{Results}\label{sec2}
\textit{IRIS} observes chromospheric and TR plasma in UV at high spatial ($\sim$ 0.3") and temporal resolution ($\sim 1$~s) with both slit-jaw imaging (SJI) and high-resolution spectroscopy ($\Delta \lambda =12.98\,\mathrm{m\AA}$ in FUV1: $1332\text{-}1358\,\mathrm{\AA}$; $\Delta \lambda =12.72\,\mathrm{m\AA}$ in FUV2: $1389\text{-}1407\,\mathrm{\AA}$; $\Delta \lambda =25.46\,\mathrm{m\AA}$ in NUV: $2783\text{-}2834\,\mathrm{\AA}$). 
IRIS also allows an accurate calibration of the absolute wavelength, thanks to the presence of chromospheric neutral lines in its spectral range \citep{DePontieu2014}.
Here we analyze a full-disk mosaic of \textit{IRIS} observations obtained on 2014-03-17 around the maximum of Solar Cycle 24. 
In each raster, the spectrograph slit ($0.33" \times 175"$) is advanced in the direction perpendicular to the slit for 64 steps, separated by 2", with a full raster covering a field of view of about $127" \times 175"$.  The (2 pixel) spatial resolution along the slit is $0.33"$, and the exposure time is 8~s for each step. Standard calibration were applied (including dark current, flat-field and geometric correction). We obtained the SDO/AIA "pseudo-mosaics" made from composition of several SDO images from https://iris.lmsal.com/mosaic\_index.html. These maps were taken at times close to the time when IRIS was sampling each portion of the disk, and were coaligned with the IRIS mosaic. 
Data potentially affected by excessive radiation spikes or other types of compromised pixels were masked out, as described in detail in \cite{Ayres2021}.

The \siiv\ 1393.76\AA\ transition region line was fitted, as thoroughly discussed in \cite{Ayres2021}, with a pseudo-Gaussian function $\Phi(\Delta\lambda) = I_{C} \exp{-(|\Delta\lambda|/\Delta\lambda_D)^\alpha}$, where $\Delta\lambda = \lambda - \lambda_0$ is the wavelength displacement relative to the line center $\lambda_0$, $\Delta\lambda_D$ is the line e-folding width, and $I_{C}$ is the central intensity $I(\lambda=\lambda_0)$. The line shape depends on the parameter $\alpha$, with $\alpha = 2$ corresponding to a Gaussian profiles, $\alpha < 2$ a more sharply peaked profile with broad wings, and $\alpha > 2$ to a more rectangular profile with truncated wings.
The pseudo-Gaussian profile for $\alpha < 2$ is similar to $\kappa$ profiles, in particular, for instance, $\alpha = 1.85$ is a close match for a $\kappa$ profile with $\kappa \sim 10$, which is relatively close to Gaussian, $\alpha = 1.65$ (which is the $\alpha$ value for the full-disk spectrum of the observation analyzed here; see Table~2 of \citealt{Ayres2021}), corresponds to $\kappa \lesssim 5$, which is distinctly non-Gaussian, and $\alpha < 1.5$ corresponds to $\kappa < 4$, i.e., strongly non-Gaussian profiles.

The full-disk maps obtained for the total line intensity and the $\alpha$ parameter (Figure~\ref{fig1}) show the presence of several active regions in the disk, and that the spatial distribution of the properties of the line profiles presents spatial coherence, particularly in active regions. 
The non-Gaussian profiles (here defined as profiles for which the $\alpha$ parameter is $<1.85$) represent about 60\% of all profiles (we mask out all pixels with \siiv\ intensity below 100~DN; see also Figure~\ref{fig4}). The intensity shows an enhancement near the limb, consistent with
an increasing plasma column depth related to line-of-sight effects, in agreement with results of previous center-to-limb studies of transition region emission \cite[e.g.,][]{Dere1984,Peter1999,McIntoshBDP2008,Kayshap2023}. The $\alpha$ map points to a limb effect, with profiles closer to Gaussian at the limb; just as for the intensity, this is also compatible with increased superposition of multiple velocity components which produces a more Gaussian shape.

We use coordinated observations with the Solar Dynamics Observatory (SDO, \citealt{Pesnell2012}, including the Atmospheric Imaging Assembly, AIA, \citealt{lemen2012atmospheric}, and the Helioseismic and Magnetic Imager, HMI, \citealt{Scherrer2012}) in different passbands sensitive to chomospheric, transition region, and coronal emission to select regions corresponding to different solar features, such as, plage (bright chromospheric emission associated with concentrated magnetic fields in solar active regions), and mossy plage (bright, structured EUV emission from the upper transition region at the footpoints of hot coronal loops in active regions).  The \textit{plage} and \textit{moss} regions were selected using SDO/AIA data (in particular 1700\AA\ and 304\AA\ for \textit{plage}, and 1700\AA, 304\AA, and 193\AA\ for \textit{moss}), following the approach of \cite{Graham2019}, and \cite{Cho2023}.
An example of the maps obtained is shown in Figure~\ref{fig2} for the active regions in the SW quadrant.

\begin{figure}[t!]
\centering
\includegraphics[width=1\textwidth]{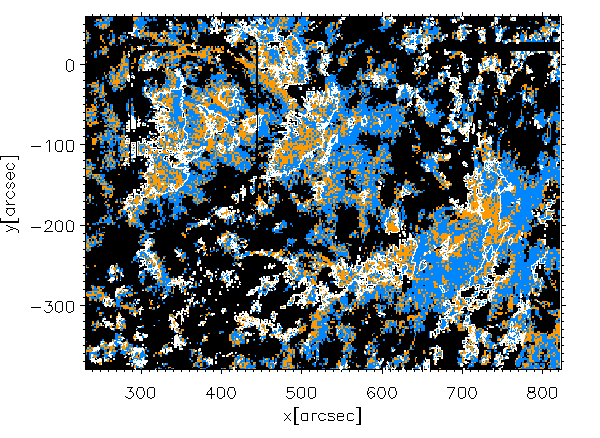}
\caption{\textbf{Map of spatial distribution of non-Gaussian/Gaussian transition region line profiles and relation with angle between magnetic field and line-of-sight.} The colored map shows $\kappa$-like ($\alpha < 1.85$) profiles in blue, and Gaussian ($1.85 < \alpha < 2.15$) profiles in orange, for pixels with total intensity above 500~DN (to reduce the potential impact of S/N on the line profile determination), for the field of view shown in Figure~\ref{fig2}. The white contours delineate regions where $\cos(i)$ is above 0.7, i.e., where the line-of-sight is more aligned with the magnetic field direction (the angle $i$ between the two is $\lesssim 45^\circ$). }\label{fig3}
\end{figure}

For optically thin transition region emission, spectroscopic diagnostics depend critically on geometry as well as local plasma conditions. Because observed Doppler shifts and line widths reflect line-of-sight velocities, the inclination angle, $i$, between the magnetic field (which largely guides plasma motions) and the observer's line of sight plays a central role in mapping physical motions onto observed spectral signatures. We use non-linear force-free field (NLFFF) extrapolation based on photospheric magnetograms from the Helioseismic and Magnetic Imager (HMI) onboard the Solar Dynamics Observatory (SDO) to derive the magnetic field vector at transition region heights. The magnetic field inclination is calculated by applying non-linear force-free field (NLFFF) extrapolations to SDO/HMI vector magnetograms obtained at a similar time as the IRIS observations. We use the NLFFF package in IDL SolarSoft for the NLFFF magnetic field extrapolation \citep{Wheatland2000} as described in detail in \cite{Cho2024}. We then assume a transition region height of $\sim 2000$~km to derive the angle $i$ (used in  Figures~\ref{fig2}, \ref{fig3}, \ref{fig4}) between the LOS and the magnetic field orientation at the transition region.
Figure~\ref{fig2} shows the map of $\cos(i)$ for the active regions in the SW quadrant of the solar disk, together with the corresponding maps of IRIS \siiv\ and AIA images.

The comparison of the maps for $\alpha$ and $\cos(i)$ clearly points to a prevalence of Gaussian profiles for larger $\cos(i)$ values (i.e., line-of-sight, LOS, more aligned with the magnetic field direction at transition region heights) and of $\kappa$-like profiles for smaller $\cos(i)$ values (larger angle between LOS and magnetic field orientation). This is shown clearly in Figure~\ref{fig3} (as well as in Figure~\ref{fig4}) where we impose an intensity threshold (500~DN) to minimize the possible impact of noise in the line profile fit, and use a simple classification of the \siiv\ profiles as non-Gaussian or Gaussian ($\alpha < 1.85$, and $1.85<\alpha<2.15$ respectively; see text in Sec.~\ref{sec2} mapping $\alpha$ values to $\kappa$ values). Figure~\ref{fig4} shows the distributions of the Gaussian/non-Gaussian \siiv\ line profiles as a function of the intensity of the \siiv \ line and of $\cos(i)$. The histograms clearly show how $\kappa$-like profiles are dominant when the LOS is more perpendicular to the magnetic field direction.

The relation we observe between the line profiles and the viewing angle indicates that the observed velocity distributions in the solar transition region are anisotropic and non-Maxwellian, and their observed properties depend systematically on the viewing angle relative to the magnetic field direction.

\begin{figure}[h]
\centering
\includegraphics[width=1\textwidth]{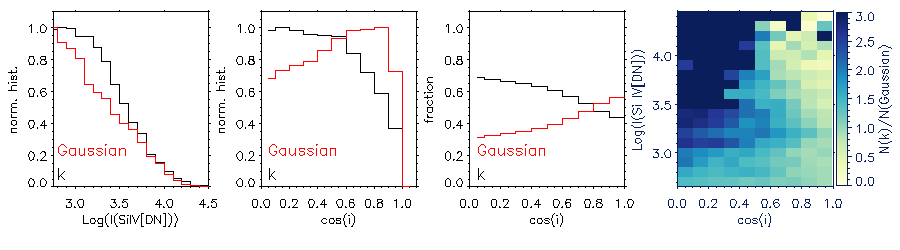}
\caption{
\textbf{Distributions and relative weight of types of \siiv\ line profiles as a function of \siiv\ line intensity and $\cos(i)$.} 
The first two histograms show distributions (each normalized to its peak) of Gaussians and $\kappa$-like profiles as a function of \siiv\ intensity and $\cos(i)$. The third plot show the fraction of Gaussians and $\kappa$-like profiles in each $\cos(i)$ bin. The last plot shows a 2D histogram of the ratio of $\kappa$-like vs.\ Gaussians as a function of \siiv\ intensity and $\cos(i)$.
All plots are for the field of view shown in Figure~\ref{fig2}.}\label{fig4}
\end{figure}

\begin{figure}[h]
\centering
\includegraphics[width=1\textwidth]{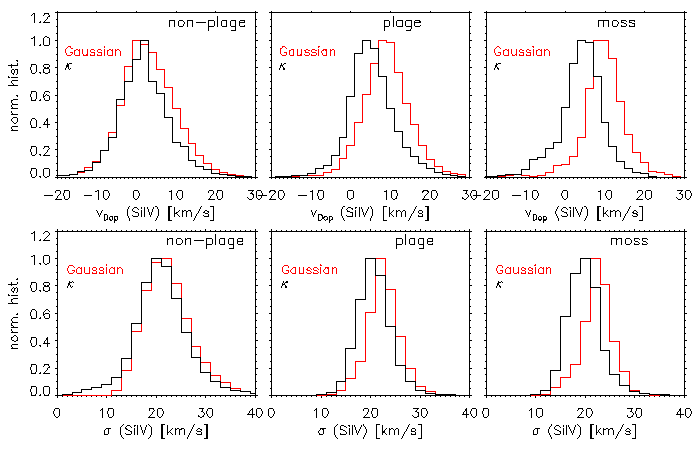}
\caption{\textbf{Distributions of Gaussian profiles and $\kappa$-like profiles as a function of line shift and width, for different solar features: non-plage, plage and moss.}
We plot histograms (each normalized to its peak) for Gaussian and $\kappa$-like profiles vs.\ line shift (top row) and line width (bottom row) in different solar features: non-plage (left column), plage (center), moss (right column). The criteria used to select the solar features are described in detail in Sec.~\ref{sec2}. The full-disk data have been used to derive these histograms.
}\label{fig4a}
\end{figure}

The Gaussian and non-Gaussian profiles also show inherent differences in \siiv\ Doppler shift and line width: non-Gaussian profiles appear narrower and less red-shifted than non-Gaussian profiles, especially in plage and in moss (Figure~\ref{fig4a}; the plotted $\sigma$ is the measured $\Delta\lambda_D$ of the pseudo-Gaussian function, and it relates to the FWHM shown in \citealt{Ayres2021} as $\sigma = \Delta\lambda_D = FWHM /(2 \ln{2}^{(1/\alpha)})$). These results point to genuinely distinct properties of the non-Gaussian and Gaussian profiles, in the more magnetically active locations. 
The Doppler shift distributions are compatible with the interpretation of $\kappa$-like profiles being caused by supra-thermal particles, as previous studies have found less red-shifted distribution of \siiv\ for small heating events with accelerated electrons \citep{Testa2014,Polito2018,Testa2020,Testa2023b,Cho2023,Cho2026}. As for the width, significant superposition of emission profiles from different substructures with a wide range of LOS velocities would simultaneously lead to larger widths and more Gaussian profiles, depending on the exact underlying LOS velocity distribution. Our observations of the line width and non-Gaussianity provide constraints on the properties of the underlying LOS velocity distribution that would be required in this scenario. 

\section{Discussion}\label{sec12}
High resolution spectral observations of transition region spectral lines capture the signatures of energy deposition and redistribution at the boundary between the lower and upper atmosphere, and have significant relevance for investigating the physical processes responsible for heating the solar atmosphere.
While the non-Gaussianity of transition region emission lines in some locations was established long ago (since the early spectroscopic observations with \textit{Skylab}, and following observations with HRTS and SOHO; e.g., \citealt{Kjeldseth1977,Doschek1978,Dere1993,Chae1998}), a satisfactory explanation remained elusive.
The mystery of the origin of the broad wings in transition region lines profiles is also relevant to stellar astrophysics as non-Gaussian profiles are also observed in stars \citep{Wood1997,Pagano2004,Linsky2012,Ayres2015}. Additionally, non-Gaussian profiles have recently been reported in solar flares \citep{Jeffrey2016,Jeffrey2017,Polito2018_kappa}, although the number of reported detections is currently small.

Using high-resolution IRIS data here we find that non-Gaussian profiles are exceedingly more common than previously observed, and constitute $\sim 60$\% of the profiles.
The widespread occurrence of non-Gaussian profiles we observe at the highest spatial resolution with \textit{IRIS} challenges previous interpretations of the non-Gaussianity as a superposition of narrow and broad dynamical components \citep[e.g.,][]{Peter2000}, favoring instead alternative formation mechanisms, such as those involving $\kappa$ distributions \citep[e.g.,][]{Dudik2014,Dudik2017,Dudik2017rev,Jeffrey2016,Jeffrey2017,Polito2018_kappa}.
Here we combine \textit{IRIS} spectral observations with magnetic field extrapolations and find a distinctive distribution of more prevalent Gaussian profiles when the line-of-sight is more aligned with the magnetic field direction at transition region heights, and more $\kappa$-like distributions when the line-of-sight is more inclined with respect to the magnetic field. This provides evidence for a previously unknown observed anisotropy of non-Maxwellian velocity distributions in the solar transition region. 
A purely multi-component line-of-sight superposition, as suggested by e.g., \cite{Peter2000}, would not naturally produce a systematic dependence on magnetic field orientation, therefore this previous interpretation of the non-Gaussian profiles is difficult to reconcile with our observations.

The presence of non-Maxwellian distributions is expected as a result of magnetic reconnection events, which are thought to play a major role in powering the solar atmosphere, especially in the solar active regions.
Several scenarios may play a role in the observed anisotropy of the non-Maxwellian distributions. 
First, we consider the scenario in which the generation of such distributions is isotropic. 
For this scenario, two cases can be considered. In the first, the generation is isotropic, but becomes anisotropic because the thermalization is more efficient along the magnetic field, while more efficiently suppressed in the direction perpendicular to the field. In such a scenario, the observations would preferentially reveal non-Maxwellian profiles when observed at large angles with respect to the magnetic field direction. This scenario is also consistent with our results showing that $\kappa$-like profiles tend to have smaller Gaussian line width.
A second case is where the generation of non-Maxwellian distributions is isotropic, but the observed dependence of the spectral line profile on the angle between the LOS and magnetic field is created by superposition that somehow is anisotropic with respect to the magnetic field. An example of such a scenario is where the underlying microscopic velocity distribution of each contributing element is narrow and non-Maxwellian, and the distribution of microscopic velocities is broader along the magnetic field line (than across the field), causing a more Gaussian, and broader, resulting line profile when observed along the magnetic field direction; this could happen for instance at the footpoints of coronal loops, where in the transition region there is a large velocity gradient, as described e.g., in \cite{Cho2024}. The observed line width and degree of Gaussianity provide constraints on the LOS microscopic velocity distributions needed to reproduce them.

The second scenario is where the generation of non-Maxwellian distributions is itself anisotropic. This could occur, for example, through processes related to resonant wave-particle interactions with ion-cyclotron waves that are generated during reconnection, and that would preferentially energize ions in the direction perpendicular to the magnetic field. The latter would be compatible with recent IRIS observations of mass-dependent energization of ions by \cite{Bahauddin2021}, and would naturally explain the increased prevalence of $\kappa$-like profiles at large viewing angles relative to the magnetic field.
Perhaps a combination of all of these effects explain our observations. 
In any case, our results suggest that we are observing a subtle remnant of the dominant heating processes in the low solar atmosphere, which appear to involve kinetic processes that are not captured by magnetohydrodynamic approaches.

The anisotropic, non-Maxwellian velocity distributions commonly observed in the solar wind demonstrate that magnetized plasmas naturally develop direction-dependent structure in velocity space \citep[e.g.,][]{Marsch2006,Verscharen2019}. Although the transition region is possibly more collisional, and observed remotely, the dependence of the line shape on the LOS-magnetic-field angle suggests that a similar anisotropy in the underlying velocity distribution may be present, with different properties along and across the magnetic field. Our findings show a new diagnostics for identifying kinetic processes in plasmas in the solar atmosphere.

%% Please use the acknowledgment and contribution environments. This will 
%% be anonomyized when the "anonymous" style option is used. 
\begin{acknowledgments}
We thank the anonymous reviewer for their careful reading of our manuscript and for their valuable comments and suggestions, which have helped us improve the paper.
P.T. was supported for this work by contract 8100002705 (IRIS), and by contract SP02H1701R (AIA) to the Smithsonian Astrophysical Observatory.  B.D.P.\ and K.C.\ were supported by NASA contract NNG09FA40C (IRIS). A.J.B.R. acknowledges support from STFC grant ST/W001195/1. IRIS is a NASA small explorer mission developed and operated by LMSAL with mission operations executed at NASA Ames Research Center and major contributions to downlink communications funded by ESA and the Norwegian Space Agency (NOSA). \end{acknowledgments}

%\begin{contribution}
%%This section gives authors the space to recognize author contributions. The text inside this environment is NOT counted towards the total word quanta. At a minimum, manuscripts are expected to include this text:

%All authors contributed equally to the Terra Mater collaboration.

%% But authors are expected to provide more specific details, e.g. 
%%
%%SC was responsible for writing and submitting the manuscript.
%%WWM came up with the initial research concept and edited the manuscript.
%%OTS obtained the funding and edited the manuscript.
%%EBF provided the formal analysis and validation. He also edited the manuscript.
%%GEH Supervised the undergraduates, wrote the software and administers the project github and Zenodo repositories.
%%
%% Authors can use the Contributor Role Taxonomy (CRediT) at
%% for ideas on how write a good statement tailored to their needs.

%\end{contribution}

\facilities{IRIS, SDO}

%\software{astropy \citep{2013A&A...558A..33A,2018AJ....156..123A,2022ApJ...935..167A},           Cloudy \citep{2013RMxAA..49..137F},      Source Extractor \citep{1996A&AS..117..393B}}


%% For this sample we use BibTeX plus aasjournalv7.bst to generate the

\bibliography{sample701}{}
\bibliographystyle{aasjournalv7}



\end{document}